# Superconductivity in Se-doped $La_2O_2Bi_2Pb_2S_{6-x}Se_x$ with a $Bi_2Pb_2Ch_4$-type thick conducting layer

Rajveer Jha[1], Yosuke Goto[1], Tatsuma D. Matsuda[1], Yuji Aoki[1], and Yoshikazu Mizuguchi[1*]

[1]*Department of Physics, Tokyo Metropolitan University, 1-1 Minami-Osawa, Hachioji, Tokyo 192-0397, Japan.*

**Abstract**

$La_2O_2Bi_2Pb_2S_6$ is a layered Bi-based oxychalcogenide with a thick four-layer-type conducting layer. Although $La_2O_2Bi_2Pb_2S_6$ is a structural analogue of $La_2O_2Bi_3AgS_6$, which is a superconductor, insulating behavior has been observed in $La_2O_2Bi_2Pb_2S_6$ at low temperatures, and no superconductivity has been reported. Herein, we demonstrate superconductivity in $La_2O_2Bi_2Pb_2S_{6-x}Se_x$ via partial substitution of Se in the S sites. Owing to the Se doping, the normal state electrical resistivity of $La_2O_2Bi_2Pb_2S_{6-x}Se_x$ at low temperatures was dramatically suppressed, and superconductivity was observed at a transition temperature ($T_c$) of 1.15 K for $x = 0.5$. $T_c$ increased with increasing Se concentration: $T_c = 1.9$ K for $x = 1.0$. The emergence of metallicity and superconductivity was explained via in-plane chemical pressure effects that can suppress local disorder and carrier localization, which are commonly observed in two-layer-type $BiS_2$-based systems.

---

[*] Corresponding author: mizugu@tmu.ac.jp

# I. INTRODUCTION

Layered compounds have been widely studied in the search for new superconductors since the discovery of (high-transition-temperature) high-$T_c$ cuprates [1]. After the cuprates, Fe-based and BiCh$_2$-based superconductors (Ch: S and Se) have been discovered, and many others have been synthesized [2-5]. In those layered superconductors, the important structure is a superconducting layer commonly contained in the layer stacking: a CuO$_2$ layer in cuprates and an FeAs or FeSe layer in Fe-based superconductors [6-8]. In BiCh$_2$-based superconductors, a BiCh$_2$ layer is the basic superconducting layer [3-5].

Since the discovery of BiS$_2$-based superconductors, namely, Bi$_4$O$_4$S$_3$ and La(O,F)BiS$_2$, in 2012 [3,4], many related superconductors with BiS$_2$[9-18], BiSe$_2$ [19,20], or Bi(S,Se)$_2$ superconducting layers [21-24] have been synthesized. The crystal structure of BiCh$_2$-based compounds has an interesting feature, that is, the presence of a van der Waals gap between two BiCh$_2$ layers. The space in this gap could be used to create new BiCh$_2$-type compounds with a thick conducting layer [25-27].

As studied in cuprates, stacking many conducting planes along the $c$-axis could be a promising way to achieve a higher $T_c$ in a layered system [6,28,29]. Hence, layered oxychalcogenides, such as La$_2$O$_2$M$_4$S$_6$ (M: metals like Bi, Pb, Sn, and Ag), with a thicker *four-layer-type* conducting layer of M$_4$S$_6$ have been synthesized with the expectation of demonstrating a higher $T_c$ than that observed in two-layer-type BiCh$_2$-based superconductors [25-27]. The four-layer-type La$_2$O$_2$M$_4$S$_6$ compounds typically have a tetragonal structure with a space group of $P4/nmm$. The structure can be regarded as a stacking of La$_2$O$_2$, BiS$_2$, M$_2$S$_2$, and BiS$_2$ layers in sequence [26]; a crystal structure image for M = Pb is displayed in Fig. 1(d). A rock-salt-type M$_2$S$_2$ layer is inserted between two BiS$_2$ layers where a van der Waals gap would exist in a two-layer-type LaOBiS$_2$ (described as La$_2$O$_2$Bi$_2$S$_4$ for one unit cell). In fact, the outer layers are basically BiS$_2$ layers, and the inner rock-salt layer is occupied by M and S ions. From a structural viewpoint, the composition can be described as La$_2$O$_2$Bi$_2$M$_2$S$_6$.

Recently, we observed superconductivity with a $T_c$ of 0.5 K in La$_2$O$_2$Bi$_3$AgS$_6$ (M = Bi$_{0.5}$Ag$_{0.5}$) [30]. Furthermore, we have studied the effects of doping on the superconductivity of the La$_2$O$_2$Bi$_3$AgS$_6$ system. Partial substitution of Ag by Sn increased $T_c$ up to 2.5 K in La$_2$O$_2$Bi$_3$Ag$_{0.6}$Sn$_{0.4}$S$_6$ [31]. Additionally, partial Se substitution for S resulted in bulk superconductivity at 3.5 K in La$_2$O$_2$Bi$_3$Ag$_{0.6}$Sn$_{0.4}$S$_{5.7}$Se$_{0.3}$ [31]. $T_c$ further increased up to 4.0 K with RE-site substitution in La$_{2-x}$RE$_x$O$_2$Bi$_3$Ag$_{0.6}$Sn$_{0.4}$S$_6$ (RE = Eu, Nd, and Pr) compounds [32,33]. Based on these recent developments with regard to new four-layer-type Bi-based superconductors, we found that the chemical pressure effect was useful for inducing bulk superconductivity in La$_2$O$_2$Bi$_2$M$_2$S$_6$ systems. The chemical pressure effect has been proposed as one of the essential parameters for the emergence of bulk superconductivity in typical BiCh$_2$-based compounds [14,21-24,34]. In the RE(O,F)BiCh$_2$ system, in-plane chemical pressure, which was tuned by RE-site and/or Ch-site substitutions, suppresses in-plane local disorder, which is unfavorable for the emergence of bulk superconductivity [35-39]. As mentioned previously, four-layer-type La$_2$O$_2$Bi$_2$M$_2$S$_6$ compounds are also sensitive to chemical pressure effects. Therefore, the chemical pressure could be useful for further exploration of new four-layer-type superconductors.

In this study, we have investigated the effects of substituting Se in the S sites in La$_2$O$_2$Bi$_2$Pb$_2$S$_6$, which was the first four-layer-type compound synthesized by Sun et al. and proposed as a promising thermoelectric material [25]. Although theoretical calculations suggested that La$_2$O$_2$Bi$_2$Pb$_2$S$_6$ is a metal [26], the obtained samples showed insulating behavior at low temperatures [25]. We have investigated the effects of partial substitutions of the Pb sites by Ag, Cd, In, Sn, and Sb, but all of the examined samples showed insulating characteristics at low temperatures. Therefore, in this study, we have investigated the effects of substituting Se in the S sites in La$_2$O$_2$Bi$_2$Pb$_2$S$_{6-x}$Se$_x$. Owing to the Se substitution, the insulating behavior was suppressed, and electrical resistivity that was apparently lower than that of pure La$_2$O$_2$Bi$_2$Pb$_2$S$_6$ was observed even at the lowest temperature of our system ($T = 0.4$ K). We observed superconductivity at $T_c = 1.15$ K for $x = 0.5$ and $T_c = 1.9$ K for $x = 1.0$.

## II. EXPERIMENTAL DETAILS

Polycrystalline samples of $La_2O_2Bi_2Pb_2S_{6-x}Se_x$ ($x = 0$, 0.5, and 1.0) were prepared by a solid-state reaction method. Powders of $La_2S_3$ (99.99%), $Bi_2O_3$ (99.9%), Pb (99.99%), and grains of Bi (99.999%), S (99.99%), and $Bi_2Se_3$ (pre-synthesized from Bi and Se) with a nominal composition of $La_2O_2Bi_2Pb_2S_{6-x}Se_x$ ($x = 0$, 0.5, and 1.0) were mixed using a pestle and a mortar, pelletized, sealed in an evacuated quartz tube, and heated at 690 °C for 15 h. The heated samples were reground, pelletized, and heated at 740 °C for 35 h. The optimal annealing conditions and the phase purity of the prepared samples were examined using laboratory X-ray diffraction (XRD) with a Cu-K$_\alpha$ radiation by a conventional $\theta$-$2\theta$ method. To discuss the lattice structure, Rietveld analyses were performed with RIETAN-FP [40]. We assumed perfect site selectivity of Bi and Pb based on the previous structural analyses for $La_2O_2Bi_2Pb_2S_6$ [26]. As the obtained XRD intensity and profile were not sufficient for the determination of displacement parameters, we have performed refinements with fixed isotropic displacement parameters. A schematic image of the crystal structure was drawn using VESTA [41]. The actual composition of the synthesized samples was investigated by energy dispersive X-ray spectroscopy (EDX) on a scanning electron microscope SEM [TM-3030 (Hitachi)]. The electrical resistivity down to $T = 0.5$ K was measured by a four-probe method using a $^3$He probe platform on the Physical Property measurement system (PPMS: Quantum Design). The Seebeck coefficient was measured by a four-probe method at room temperature on ZEM-3 (advance RIKO). The temperature dependence of the Hall coefficient was measured with a standard five-prove configuration on PPMS.

## III. RESULTS AND DISCUSSION

Figure 1(a) shows room temperature powder XRD patterns for $La_2O_2Bi_2Pb_2S_{6-x}Se_x$ ($x = 0$, 0.5, and 1.0) compounds. The obtained samples are crystallized in a tetragonal crystal structure with a space group of *P*4/*nmm*. For all the samples, the main phase was $La_2O_2Bi_2Pb_2S_{6-x}Se_x$. As indicated by the symbol, a $La_2O_2S$ impurity phase was observed. The impurity phase did not disappear when heat treatment conditions were changed. We performed two-phase Rietveld analyses to determine lattice constants for all the obtained samples. The results of the Rietveld refinement are summarized in Table I. The lattice constants *a* and *c* increase with Se substitution for $x = 0.5$, but the lattice constants for $x = 1.0$ were comparable to those for $x = 0.5$. Those shifts in lattice constants are consistent with the observed peak shifts, as shown in Fig. 1(b,c). Based on such facts, we consider the solubility limit of Se in S to be less than $x = 1.0$. We have analyzed the actual compositions of elements at the conducting layer in $La_2O_2Bi_2Pb_2S_{6-x}Se_x$ ($x = 0$, 0.5 and 1.0) by EDX analysis at five analysis points (Table I). The obtained compositions for the sample are almost the same as the initial nominal compositions, but the Se concentration for $x = 1.0$ is slightly lower than the nominal value. However, the Se concentrations were different for the samples with x = 0.5 and 1.0, and the superconducting property was also different. Therefore, for clarity to the readers, we labelled the samples according to the initial nominal compositions in this paper.

Figure 1(e) shows the result of Rietveld refinement for $x = 0.5$. To refine the XRD pattern for an Se-doped sample, we needed to determine the S site where Se is doped. The inset of Fig. 1(e) shows the reliability factor $R_{wp}$ obtained for the cases with no Se and those with Se-doped S1, S2, and S3 sites [see Fig. 1(d) for the label of S sites]. $R_{wp}$ decreases with incorporation of Se at the S sites and becomes the lowest when Se substitution at the in-plane S1 site was assumed. Although we need further detailed structural parameters to draw firm conclusions regarding the Se-doping site, herein, we consider that Se was selectively substituted at the S1 site. This assumption is consistent with the emergence of metallic conductivity and superconductivity in Se-doped samples, which can be explained by the concept of in-plane chemical pressure via Se substitution at the S1 sites of $BiS_2$ layers [21-24,34].

TABLE I. Rietveld analysis results, lattice constants, reliability factor $R_{wp}$, and mass fraction of the major phase to the impurity phase in $La_2O_2Bi_2Pb_2S_{6-x}Se_x$ ($x = 0$, 0.5, and 1.0).

| Sample label | $La_2O_2Bi_2Pb_2S_6$ ($x = 0$) | $La_2O_2Bi_2Pb_2S_{5.5}Se_{0.5}$ ($x = 0.5$) | $La_2O_2Bi_2Pb_2S_{5.0}Se_{1.0}$ ($x = 1.0$) |
|---|---|---|---|
| $a$ (Å) | 4.0920(4) | 4.1039 (3) | 4.1038 (4) |
| $c$ (Å) | 19.680(2) | 19.7642 (14) | 19.777 (2) |
| $R_{wp}$ | 12.8% | 10.5% | 11.7% |
| Mass fraction of the main phase | 95% | 95% | 82% |
| Actual composition (EDX) | $Bi_{1.98(3)}Pb_{2.09(8)}S_{6.04(9)}$ | $Bi_{2.01(3)}Pb_{1.89(5)}S_{5.52(8)}Se_{0.46(3)}$ | $Bi_{2.04(1)}Pb_{1.95(2)}S_{5.18(5)}Se_{0.81(10)}$ |

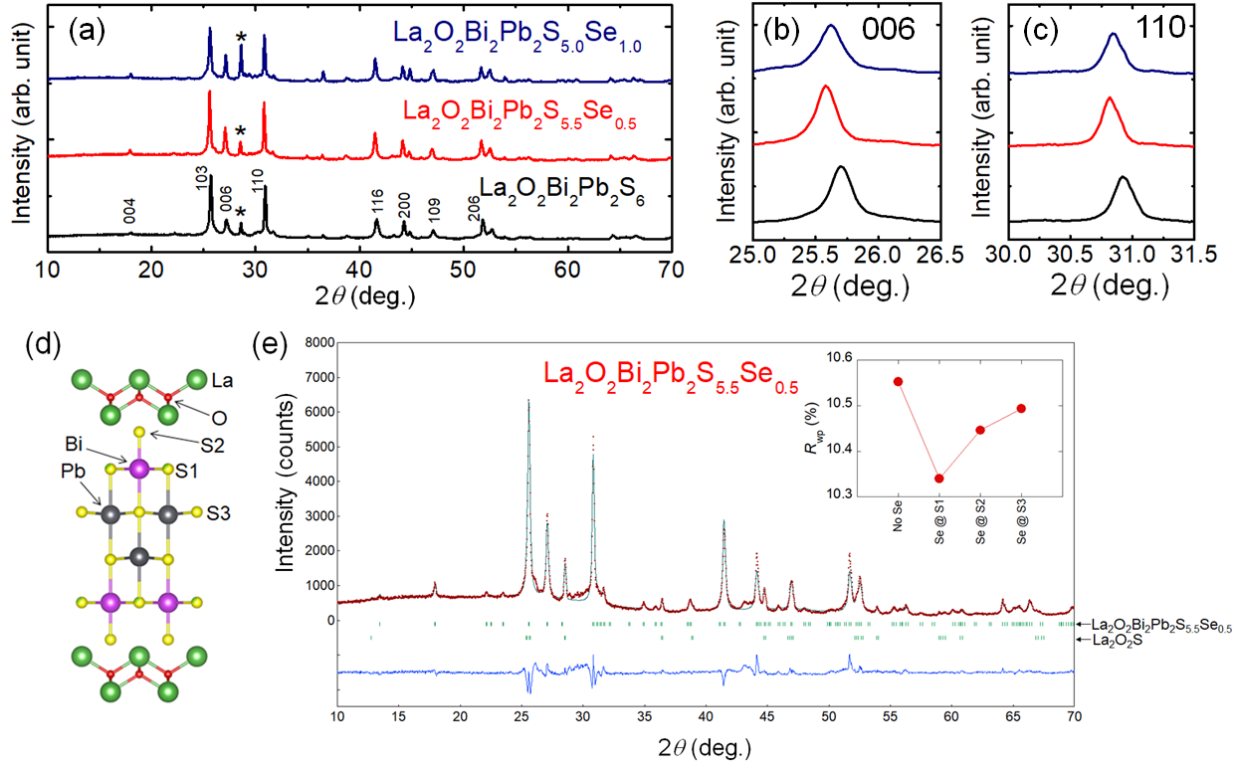

**FIG. 1.** (color online) (a) Powder XRD patterns for all the samples of $La_2O_2Bi_2Pb_2S_{6-x}Se_x$. The peak of the $La_2O_2S$ impurity phase was indicated by an asterisk symbol. The numbers in the figures are Miller indices of typical peaks. (b) Zoomed profiles of 006 peaks. (c) Zoomed profiles of 110 peaks. (d) Schematic image of $La_2O_2Bi_2Pb_2S_{5.5}Se_{0.5}$ crystal structure drawn with parameters obtained from the Rietveld analysis. (e) Powder XRD pattern and the Rietveld fitting for the $La_2O_2Bi_2Pb_2S_{5.5}Se_{0.5}$ sample. The Rietveld analysis was performed by two-phase refinement with a $La_2O_2S$ impurity phase as a second phase.

Figure 2 shows the temperature dependence of the electrical resistivity [$\rho(T)$] from 300 to 2 K for $La_2O_2Bi_2Pb_2S_6$. The electrical resistivity largely increases with decreasing temperature, which in agreement with the typical behavior of semiconductors. The result is consistent with a previous report by Sun et al. [25]. However, the band calculation has suggested a metallic band structure for $La_2O_2Bi_2Pb_2S_6$ [26]. The inconsistency may imply that the semiconducting-like behavior of electrical resistivity is caused by anomalous carrier localization, not from characteristics of a band insulator.

Figure 3(a) shows the temperature dependence of electrical resistivity [$\rho(T)$] from 300 to 0.5 K for the Se-doped $La_2O_2Bi_2Pb_2S_{6-x}Se_x$ ($x$ = 0.5 and 1.0) samples. At 300 K, the electrical resistivity is clearly lower than that of $La_2O_2Bi_2Pb_2S_6$. Furthermore, the semiconducting-like behavior was dramatically suppressed by Se substitution. The normal-state resistivity of Se-doped compounds still shows a slight increase with decreasing temperature, which may be due to weak carrier localization, which is typically observed in $BiS_2$-based systems. The weak carrier localization seems to be suppressed with increasing Se concentration from $x$ = 0.5 to $x$ = 1.0. The changes in resistivity caused by Se substitution can be understood by the effects of in-plane chemical pressure, which was introduced previously. For two-layer-type $BiCh_2$-based compounds, it has been extensively demonstrated that in-plane chemical pressure effects successfully suppressed the in-plane local disorder of the conducting layer [14,21-24,34]. Thus, we consider that the semiconducting-like behavior observed for $x$ = 0 and the weak localization behavior observed for Se-doped samples are both related to the presence of local disorders in the $BiS_2$ layers in $La_2O_2Bi_2Pb_2S_{6-x}Se_x$. We observed superconductivity in both the Se-doped samples, as shown in Fig. 3(b), which is the zoomed view of Fig. 3(a) near the superconducting transition region. $T_c$ increases with increasing Se concentration in $La_2O_2Bi_2Pb_2S_{6-x}Se_x$. The zero-resistivity $T_c$ ($T_c^{zero}$) is 1.15 K for $x = 0.5$ and 1.90 K for $x = 1.0$. Thus far, superconductivity in $La_2O_2Bi_2M_2S_6$ has been observed for $La_2O_2Bi_3AgS_6$-based compounds [30-33]. $La_2O_2Bi_3AgS_6$ exhibits metallic conductivity without Se substitution but shows a CDW-like (charge density wave-like) anomaly in $\rho(T)$. Although the origin of the anomaly has not been clarified, it could be related to the anomalous insulating behavior observed in $La_2O_2Bi_2Pb_2S_6$. The emergence of metallic conductivity and superconductivity via Se substitution in $La_2O_2Bi_2Pb_2S_{6-x}Se_x$ could provide us with important information for the further development of four-layer-type Bi-based superconductors.

Figures 4(a) and 4(b) show the temperature dependence of electrical resistivity under various magnetic fields $[\rho(T,B)]$ for $La_2O_2Bi_2Pb_2S_{6-x}Se_x$ ($x = 0.5$ and 1.0). $T_c$ shifts towards a lower temperature with an increase in the magnetic field. We observed an onset drop in the resistivity due to the superconductivity even at $B = 2$ T for $x = 1.0$. The $T_c$ onset shifts gradually with an increasing magnetic field as opposed to the shift in $T_c^{zero}$, which is the characteristic behavior of layered superconductivity against the magnetic field. The magnetic field-temperature phase diagrams are plotted in Fig. 4(c). The upper critical fields are estimated based on the temperature where resistivity becomes 90% of the normal-state resistivity near $T_c$. The upper critical field $B_{c2}(T)$ at absolute zero temperature, $B_{c2}(0)$, is analyzed using the conventional one-band Werthamer–Helfand–Hohenberg (WHH) equation [42], which gives $B_{c2}(0) = -0.693T_c(dB_{c2}/dT)_{T=T_c}$. $B_{c2}(0)$ is 0.76 and 1.15 T for $x = 0.5$ and 1.0, respectively.

To investigate the changes in carrier concentration, Seebeck coefficient ($S$) and Hall coefficient were measured for $La_2O_2Bi_2Pb_2S_{6-x}Se_x$. As shown in Fig. 5, the Seebeck coefficient is negative, which indicates that the dominant carriers are electrons. The absolute value of $S$ decreases from $x = 0$ to $x = 0.5$. Seebeck coefficient for $x = 0.5$ and 1.0 is almost the same. From the results, we can roughly estimate that the carrier concentration slightly increased by Se substitution. Fig. 6(a) shows the temperature dependence of Hall coefficient ($-R_H$) from 2 K to 300 K under the fixed magnetic field of 9 T for $La_2O_2Bi_2Pb_2S_{6-x}Se_x$ ($x = 0$ and 0.5). Hall coefficients are all negative, which is consistent with the results of Seebeck coefficient, but $-R_H$ for $x = 0.5$ is largely suppressed as compared to that for $x = 0$. To investigate the carrier density we have used the simple single band modle $n = |1/eR_H|$, where $n$ is the charge carrier density and $e$ is the electron charge, as demonstrated for other layered Bi-based compounds [25,43,44]. Figure 6 (b) shows the temperature dependences of electron carrier density for $La_2O_2Bi_2Pb_2S_{6-x}Se_x$ ($x = 0$ and 0.5), which suggests that the carriers density increases by Se substitution. Having considered the valence state of S and Se, which is considered as -2 for both, we do not expect a huge change in carrier concentration. Thus, the huge increase in carrier density may be related to the in-plane chemical pressure effects. As a fact, metalicity has been induced by Se substitution. From those facts, we consider that the huge change in carrier concentration estimated here is related to the increase in the concentration of *mobile* carrier. With this assumption, low carrier density for $x = 0$ can be considered as resulting from carrier localization due to the local disorder in the $BiS_2$ layers.

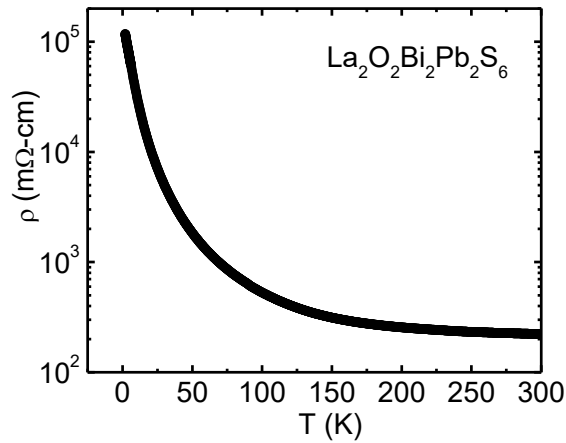

**FIG. 2. Temperature ($T$) dependence of electrical resistivity [$\rho(T)$] from 300 K to 2 K for $La_2O_2Bi_2Pb_2S_6$.**

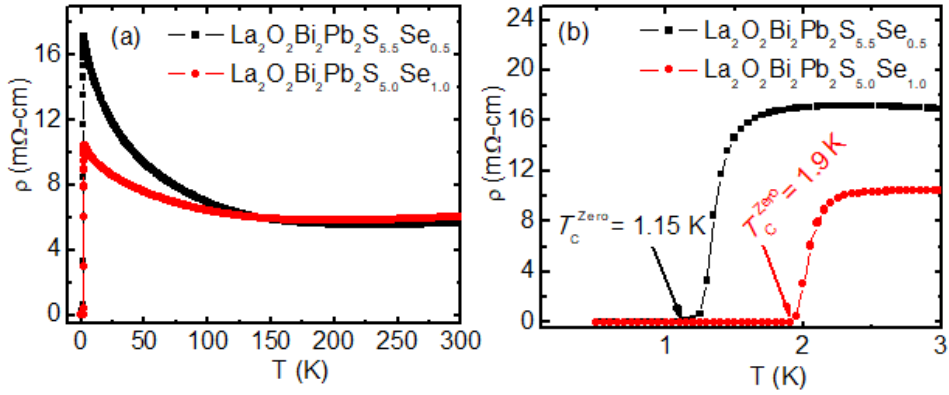

**FIG. 3. (color online) (a)** Temperature dependence of electrical resistivity from 300 to 0.5 K for La$_2$O$_2$Bi$_2$Pb$_2$S$_{6-x}$Se$_x$ ($x$ = 0.5 and 1.0). **(b)** $\rho(T)$ curve in the temperature range of 3.0-0.5 K.

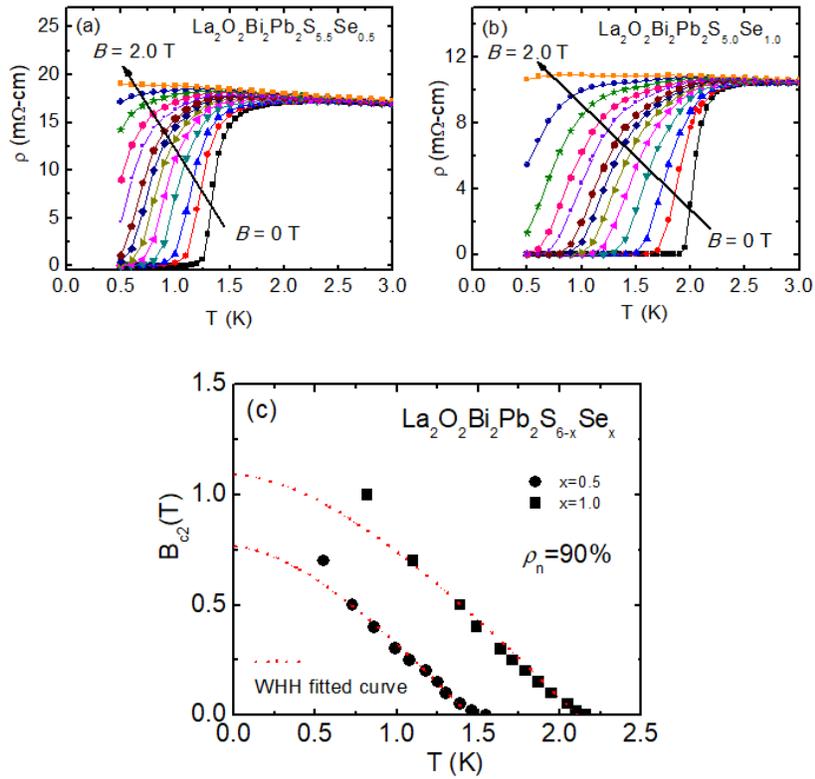

**FIG. 4. (color online) (a, b)** Temperature dependence of electrical resistivity from 3.0 to 0.5 K for La$_2$O$_2$Bi$_2$Pb$_2$S$_{6-x}$Se$_x$ ($x$ = 0.5 and 1.0) under magnetic fields. **(c)** Temperature dependence of $B_{c2}(T)$ for La$_2$O$_2$Bi$_2$Pb$_2$S$_{6-x}$Se$_x$ ($x$ = 0.5, and 1.0). The points with different symbols represent experimental data and the dotted lines represent fitting curves as per the WHH theory.

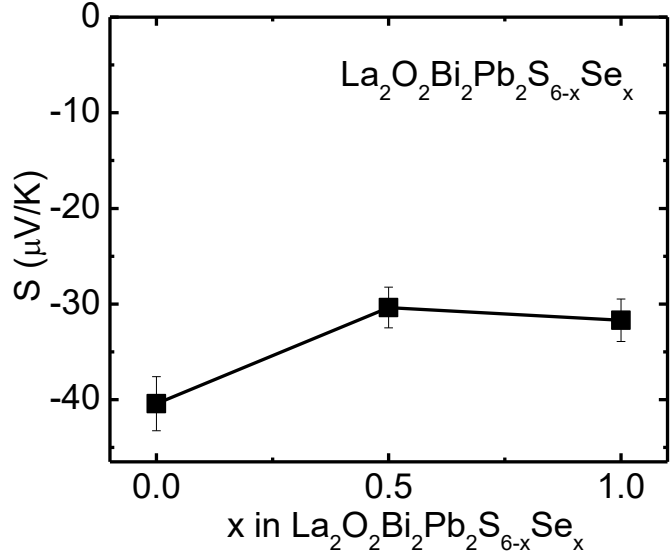

FIG. 5. The $x$ dependence of Seebeck coefficient for $La_2O_2Bi_2Pb_2S_{6-x}Se_x$ at room temperature.

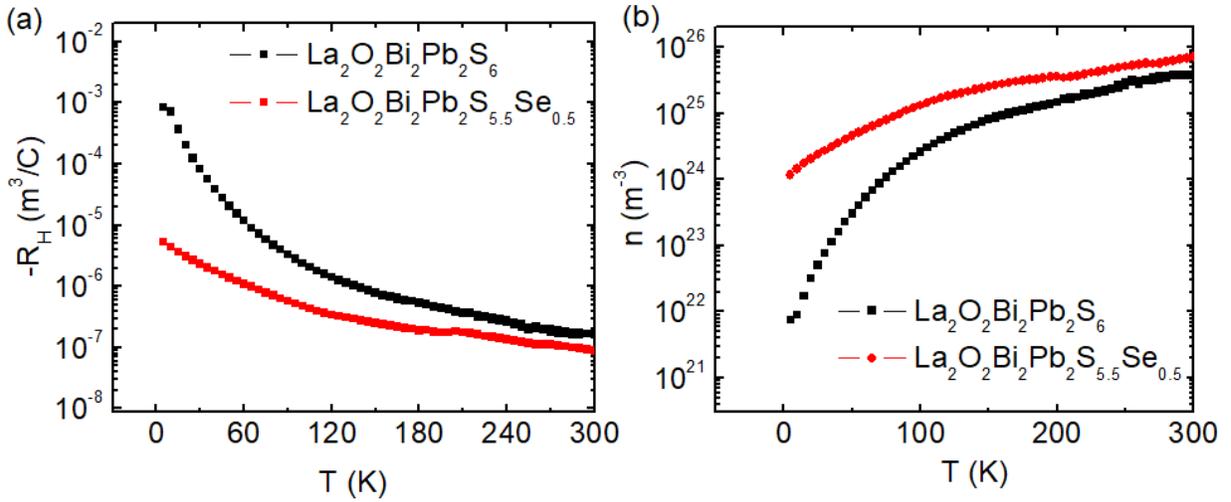

FIG 6. (a) Temperature dependence of Hall coefficient ($R_H$) for $La_2O_2Bi_2Pb_2S_{6-x}Se_x$ ($x$ = 0 and 0.5) under the magnetic field of 9 T. (b) Temperature dependence of estimated carrier concentration for $La_2O_2Bi_2Pb_2S_{6-x}Se_x$ ($x$ = 0 and 0.5).

## IV. SUMMARY

$La_2O_2Bi_2Pb_2S_6$ is a four-layer-type Bi-based compound that shows insulating transport properties at low temperatures. Hence, in this study, we have investigated the effects of Se substitution on the structural, transport, and superconducting properties of $La_2O_2Bi_2Pb_2S_{6-x}Se_x$. Owing to Se doping, the normal state electrical resistivity at low temperatures was drastically suppressed, and superconductivity at a transition temperature ($T_c$) of 1.15 K was observed for $x$ = 0.5. $T_c$ increased with increasing Se concentration: $T_c$ = 1.9 K for $x$ = 1.0. According to XRD analyses, we proposed that Se was substituted for the in-plane S1 site in the $BiS_2$ layers. Both the Seebeck coefficient and the Hall coefficient suggested that electrons were the dominant carriers. From the Hall coefficient, carrier concentration was estimated. Owing to Se substitution, the concentration of mobile carriers increased considerably; this could be related to the suppression of carrier localization caused by Se substitution. On the basis of the obtained results and the discussion based on the analogy of two-layer-type systems, the emergence of metallicity and superconductivity was explained by the in-plane chemical pressure effects. The estimated $B_{c2}(0)$ was 0.76 and 1.15 T for $x = 0.5$ and 1.0, respectively. The demonstration of metallicity and superconductivity in $La_2O_2Bi_2Pb_2S_6$-based compounds can be useful for further development of four-layer-type Bi-based superconductors.

## ACKNOWLEDGMENTS

This work was partly supported by grants in Aid for Scientific Research (KAKENHI) (Grant Nos. 15H05886, 15H05884, 16H04493, 18KK0076, and 15H03693) and the Advanced Research Program under the Human Resources Funds of Tokyo (Grant Number: H31-1).